\documentclass{osa-article}
\newcommand{\change}{}
\newcommand{\NewChange}{}

\journal{oe}
\begin{document}
	
	\title{Design and fabrication of ridge waveguide-based nanobeam cavities for on-chip single-photon sources}
	
	\author{
		U\u{g}ur Meri\c{c} G\"ur,\authormark{1,2}
		Yuhui Yang,\authormark{2}
		Johannes Schall,\authormark{2}
		Ronny Schmidt,\authormark{2}
		Arsenty Kaganskiy,\authormark{2}
		\NewChange{Yujing Wang,\authormark{3}}
		Luca Vannucci,\authormark{3}
		Michael Mattes,\authormark{1}
		Samel Arslanagi\'c,\authormark{1}
		Stephan Reitzenstein,\authormark{2}
		and Niels Gregersen\authormark{3,*}
	}
	
	\address{
		\authormark{1}Department of Electrical Engineering, Technical University of Denmark, Ørsteds Plads, Building 348, 2800 Kongens Lyngby, Denmark\\
		\authormark{2}Institut für Festkörperphysik, Technische Universität Berlin, Hardenbergstraße 36, 10587 Berlin, Germany\\
		\authormark{3}Department of Photonics Engineering, Technical University of Denmark, Ørsteds Plads, Building 343, 2800 Kongens Lyngby, Denmark}
	
	\email{\authormark{*}ngre@fotonik.dtu.dk} %% email address is required
	
	
	
	\begin{abstract*}
		We report on the design of nanohole/nanobeam cavities in ridge waveguides for on-chip, quantum-dot-based single-photon generation. Our design overcomes limitations of a low-refractive-index-contrast material platform in terms of emitter-mode coupling efficiency and yields \change{an outcoupling efficiency of 0.73 to the output ridge waveguide.} Importantly, this high coupling efficiency is combined with broadband operation of 9 nm full-width half-maximum. We provide an explicit design procedure for identifying the optimum geometrical parameters according to the developed design.
		Besides, we fabricate and optically characterize a proof-of-concept waveguide structure. The results of the microphotoluminescence measurements provide evidence for cavity-enhanced spontaneous emission from the quantum dot, thus supporting the potential of our design for on-chip single-photon sources applications.
	\end{abstract*}
	
	%%%%%%%%%%%%%%%%%%%%%%%%%%  body  %%%%%%%%%%%%%%%%%%%%%%%%%%
	\section{Introduction}\label{sec:introduction}
	Within optical quantum information processing, a key device is the single-photon source \cite{Gregersen2017} (SPS) capable of producing single indistinguishable photons on demand. Deterministic sources based on a semiconductor quantum dot (QD) \cite{Michler2017} integrated into nanophotonic structures \cite{Tomm_2021,liu2019solid,Wang2019b,Somaschi2016} have recently emerged as a leading platform for highly efficient generation of single indistinguishable photons. Such structures are suitable for coupling light into an optical fiber for use within quantum communication \cite{Takemoto2015} and for the realization of plug$\&$play SPSs \cite{schlehahn2018stand,musial2020plug}. Furthermore, boson sampling experiments \cite{Wang2018b,Wang2019c} have been performed using quantum dot single-photon sources. Nevertheless, for fully integrated quantum photonics integration of single QDs into on-chip quantum circuits \cite{arcari2014near,uppu2020scalable, Liu2018, fattah2013efficient,stepanov2015quantum,schnauber2018deterministic,mrowinski2019directional, hoehne2019numerical, hepp2020purcell, jons2015monolithic, reithmaier2013chip,rodt2021integrated,uppu2020chip}
	is required.
	
	\begin{figure}[b]
		\centering
		{\includegraphics[width=8cm]{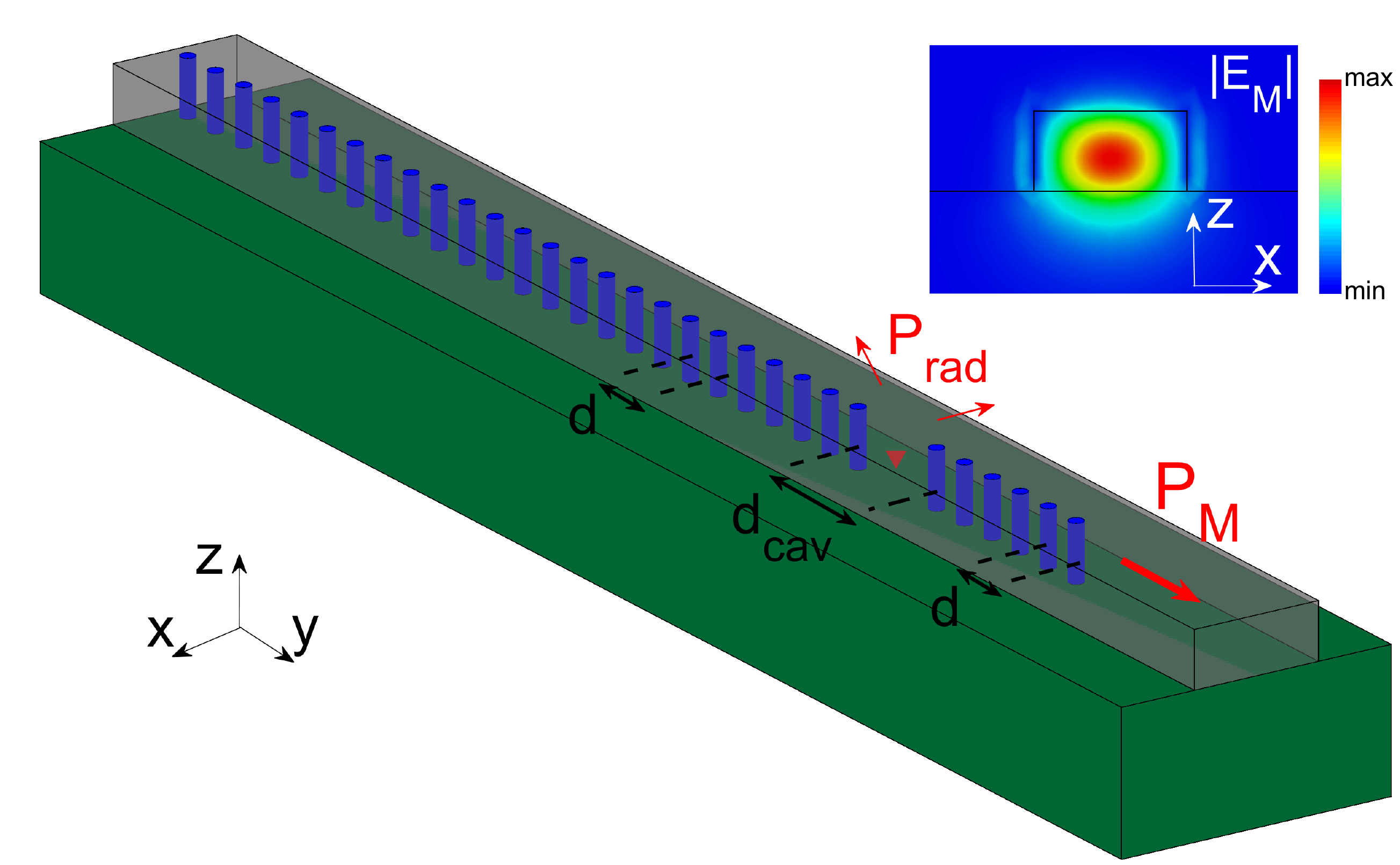}}
		\caption{Overview of the cavity ridge waveguide structure. A quantum dot (red triangle) is placed at the center of the cavity sandwiched between reflectors formed by periodic arrays of holes (blue cylinders). \change{$\rm P_{\rm M}$} denotes the power funneled into the fundamental guided mode M in the output ($+y$) waveguide, and $\rm P_{\rm rad}$ indicates the power lost to radiation modes. The inset shows the magnitude of the electric field intensity of the fundamental guided mode M.} \label{fig:intro}
	\end{figure}
	
	An attractive platform within on-chip QD-based single-photon generation is the suspended photonic crystal\cite{arcari2014near,uppu2020scalable, Liu2018} (PC) waveguide, where a spontaneous emission $\beta$ factor describing the relative QD-waveguide coupling efficiency exceeding $\sim$~0.9 has been demonstrated \cite{arcari2014near,uppu2020scalable} combined with single-photon indistinguishability of $\sim$~0.96 over long delays \cite{uppu2020scalable}. However, drawbacks of the PC waveguide approach include complex fabrication, high propagation loss \cite{Rigal2017} and in turn the need for a PC-to-ridge waveguide coupler \cite{fattah2013efficient}.
	
	An alternative design approach is the ridge waveguide \cite{stepanov2015quantum, schnauber2018deterministic,mrowinski2019directional, hoehne2019numerical, hepp2020purcell, fattah2013efficient, jons2015monolithic, reithmaier2013chip} offering mechanical robustness combined with propagation loss below 2 dB/cm  \cite{Pu2016,inoue1985low}.
	For a GaAs waveguide on a glass SiO$_2$ substrate, simply choosing the optimum waveguide width and height parameters allows for a $\beta \sim$~0.6 thanks to the large GaAs/SiO$_2$ index contrast \cite{stepanov2015quantum}. However, placing the waveguide on a III-V AlGaAs substrate significantly simplifies the fabrication process, and recent progress in deterministic integration of QDs in ridge waveguides \cite{schnauber2018deterministic,mrowinski2019directional} was based on the AlGaAs substrate platform. 
	A drawback of the GaAs/AlGaAs material system is the low index contrast leading to significant light emission into the substrate resulting in an outcoupling efficiency $\sim$~0.2 \cite{hoehne2019numerical} for a truncated waveguide.
	\change{There is thus a need for new SPS designs based on the low-index-contrast platforms with high outcoupling efficiency.}
	%\change{Clearly, overcoming the limitations of low-index-contrast platforms would enable the use of ridge waveguides --- instead of suspended structures --- for quantum photonic integrated circuits. This, in turn, would be a very important advantage for photonic quantum technologies.}
	
	The QD-waveguide coupling efficiency can be improved using engineering of the photonic environment, in particular by introducing a cavity and exploiting Purcell enhancement \cite{Liu2018,hepp2020purcell,krasnok2015antenna} to ensure preferential emission into the cavity mode. By introducing periodic arrays of holes to form distributed Bragg reflectors (DBRs) in the waveguide, so-called nanobeam cavities are formed. \change{More generally, the nanobeam design approach has allowed for the demonstration of strong coupling \cite{Ohta2011}, thresholdless lasing \cite{Jagsch2018} as well as cavity-enhanced interaction for Si vacancy centers in diamond \cite{Sipahigil2016, Nguyen2019} and for rare earth ions \cite{Raha2020}.}
	In a recent work \cite{hepp2020purcell}, a narrow-band nanobeam SPS design based on rectangular DBR holes with a \change{coupling efficiency} of 0.86 has been presented. Here, a limitation was the weak suppression of light propagation in the 400/150 DBR period configuration leading to a large mode volume V. In turn, a large cavity Q factor was needed to achieve significant Purcell enhancement, which subsequently required the implementation of a (strain-) tuning mechanism. A \textit{broadband} ridge waveguide design eliminating the need for complicated tuning mechanisms is thus desirable. 
	
	In this work, we introduce a ridge waveguide nanobeam cavity design based on the GaAs/AlGaAs material system consisting of DBRs formed by circular holes as illustrated in Fig.\ \ref{fig:intro}. \change{The introduction of an one-sided cavity leads to a relative improvement of the outcoupling efficiency} by a factor of 6.6 compared to the bare waveguide, resulting in a \change{simulated total efficiency  $\approx 0.73$} combined with broadband operation with FWHM of $\sim$ 9 nm.
	\change{We focus on embedding a quantum emitter in between the cavity mirrors and on achieving highly efficient coupling to the fundamental waveguide mode. We leave optimization of the cavity design using tapering of the mirror section \cite{Panettieri2016, Olthaus2020, Zhan2020} or using an elliptical hole shape \cite{Quan2011, Panettieri2016, Zhan2020} to a follow-up work.}
	To demonstrate the potential of our results for on-chip photonic quantum technologies, we have fabricated a large-diameter, non-optimized version of our design.
	Indeed, the results from the optical characterization strongly support the presence of cavity-enhanced spontaneous emission from this SPS device, in agreement with theoretical predictions.
	
	The manuscript is organized as follows: Section \ref{sec:Design} presents the design procedure for the ridge waveguide and the cavity structure. Section \ref{sec:fab} presents the fabricated device and the optical characterization. In Section \ref{sec:discussion}, we discuss the strengths and the weaknesses of the approach and perspectives for further improvement, followed by a conclusion in Section \ref{sec:conc}.

	\section{Design and simulations} \label{sec:Design}
	
	The numerical simulations were performed using a commercial finite element method (FEM) based solver, CST Studio Suite \cite{CST}. 
	We consider the ridge waveguide platform depicted in Fig.\ \ref{fig:intro} consisting of a GaAs waveguide placed on an $\rm Al_{0.9}Ga_{0.1}As$ substrate. The design wavelength is chosen as $\lambda_0$ = 945 nm corresponding to an InAs QD. In the following, we consider as performance figure of merit the \change{coupling efficiency, $\varepsilon$,} representing the fraction of light emitted from a QD placed in the center of the waveguide into the fundamental transverse electric (TE) waveguide mode M (c.f. Fig.\ \ref{fig:intro}, inset) in the output ($+y$) waveguide.
	We assume a QD with in-plane ($xy$) dipole orientation modeled as a classical Hertzian point dipole, and excited with off-resonant laser light as described in Section \ref{sec:fab}. The corresponding \change{efficiency $\varepsilon_{xy}$} is given by \cite{hoehne2019numerical}
	\begin{equation}
		\change{\varepsilon_{xy}} = \frac{\rm P_{\rm M,\textit x}+\rm P_{\rm M,\textit y}}{\rm P_{\rm \textit x}^{\rm T}+\rm P_{\rm \textit y}^{\rm T}}, \label{epxy}
	\end{equation}  
	where $\rm P_{\rm M}$ is the power funneled into the fundamental guided mode M in the output ($+y$) waveguide, and $\rm P^{\rm T}$ is the total power emitted from the QD including a contribution $\rm P_{\rm rad}$ lost to radiation modes.
	The subscripts $x$ and $y$ denote the dipole orientation of the emitter in the calculation of the power. 
	\change{We stress that, the figure of merit $\varepsilon_{xy}$ is obtained from the output power calculated at a location after the right mirror, that is \emph{outside the cavity}. In our case of an emitter placed inside a cavity, $\varepsilon_{xy}$ thus differs from the spontaneous emission $\beta$ factor by taking into account both the relative coupling to the cavity mode (the $\beta$ factor) and the subsequent non-unity transmission from the cavity into the output waveguide channel, parameters which are sometimes analyzed separately \cite{wang2020micropillar}.}
	$\rm P_{\rm M}$ is obtained in CST Studio Suite using waveguide ports, which project the total power in the structure onto the individual waveguide modes.
	
	We also define the \change{efficiency $\varepsilon_{x}$} for a dipole in the $x$ direction as
	\begin{equation}
		\change{\varepsilon_{x}} = \frac{\rm P_{\rm M,\textit x}}{\rm P_{\rm \textit x}^{\rm T}}. \label{epx}
	\end{equation}  
	We note that the electric field $y$ component of the TE mode is zero at the waveguide center, and thus only an $x$-polarized emitter contributes to the fundamental TE mode ($\rm P_{\rm M,\textit y}=0$). 
	In the design optimization, the initial quantity of interest is thus \change{$\varepsilon_{x}$} corresponding to the dipole $x$ orientation. However, the QD may still emit $y$-polarized light represented by the P$_{\rm \textit y}^{\rm T}$ contribution in the denominator of Eq.\ \eqref{epxy}, and the final figure of merit \change{$\varepsilon_{xy}$} will thus always be smaller than \change{$\varepsilon_{x}$}.
	Let us note that \change{$\varepsilon_{x}$} would be the relevant figure of merit if we were to address the $x$ dipole individually e.g. using phonon-assisted near-resonant excitation \cite{thomas2021bright}. However, we have characterized our device using off-resonant laser light, as an optimization of the excitation strategy is beyond the scope of this work.
	Finally, we define the generalized \cite{Gregersen2016} wavelength-dependent Purcell factor as $F(\lambda)= {\rm P_{\rm \textit x}^{\rm T}(\lambda)} / {P_{\rm bulk}(\lambda)}$, where $P_{\rm bulk}$ is the power emitted by the dipole into a bulk medium.
	
	\begin{figure}[tb]
		\centering
		{\includegraphics[width=9cm]{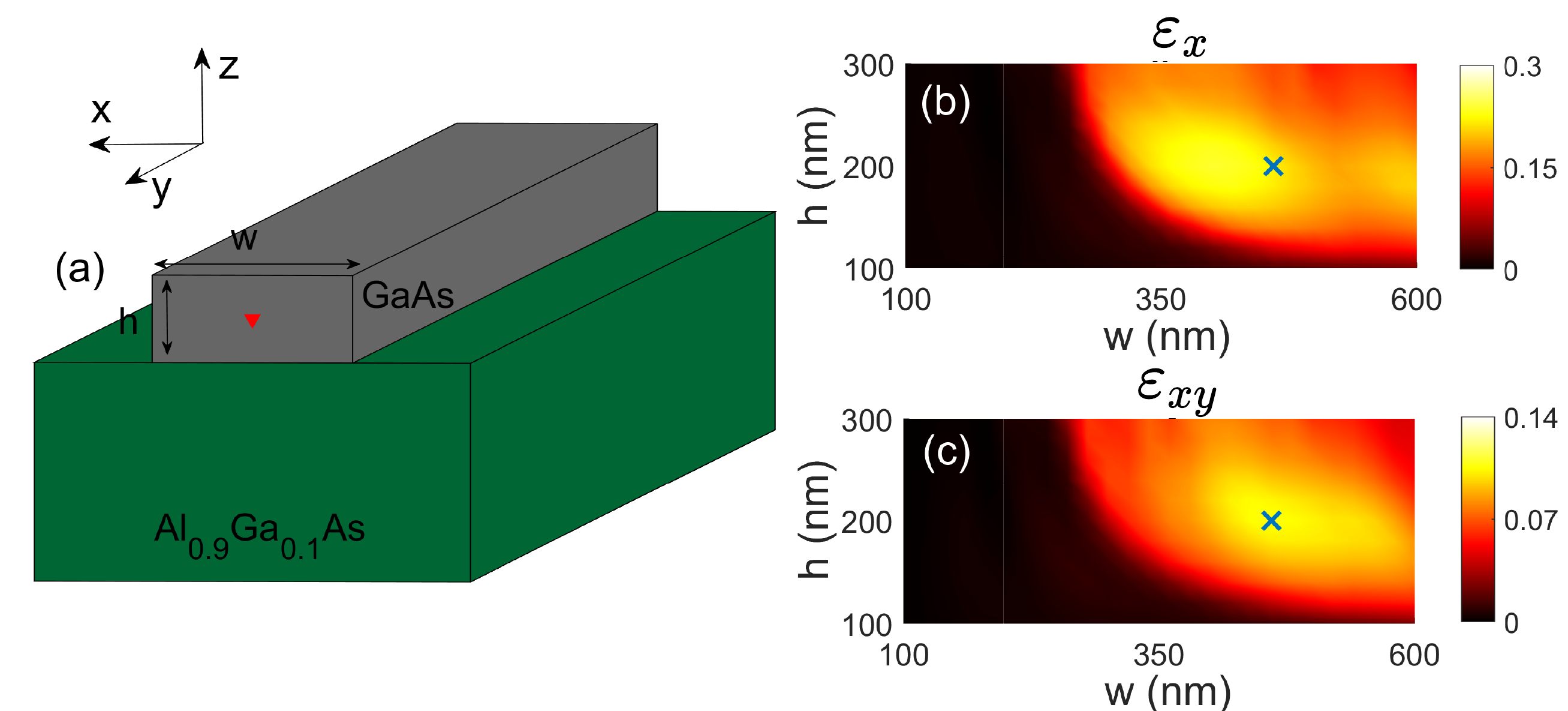}}
		\caption{(a) Bare ridge waveguide geometry. The QD is located at the center of the GaAs waveguide above the Al$_{0.9}$Ga$_{0.1}$As substrate. \change{Efficiency $\varepsilon_{x}$ (b) and  $\varepsilon_{xy}$} (c) as function of the bare waveguide width and height.} \label{fig:barewg}
	\end{figure}
	
	The starting point of the design procedure is to determine the optimum dimensions for the bare waveguide geometry without reflectors shown in Fig.\ \ref{fig:barewg}(a).  \change{The efficiency $\varepsilon_{x}$} is presented in Fig.\ \ref{fig:barewg}(b) as function of waveguide width and height, where a maximum \change{$\varepsilon_{x}$} of $\sim$ 0.24 is obtained for $h$ = 200 nm and $w$ = 420 nm.
	Now, \change{$\varepsilon_{xy}$} is shown in Fig.\ \ref{fig:barewg}(c), where the performance has been overall reduced due to the presence of the P$_{\rm \textit y}^{\rm T}$ term in Eq.\ \eqref{epxy} as discussed above. We observe that the peak value has shifted to \change{$\varepsilon_{xy} \sim$ 0.11} at $h$ = 200 nm and $w$ = 460 nm (blue crosses in Fig.\ \ref{fig:barewg}), and these optimal dimensions are used in the remainder of the design steps. 
	
	We now proceed with the design of the mirror, where we exploit that a periodic modification in the dielectric material results in a photonic band-gap allowing for high reflectivity at the center of the band-gap \cite{joannopoulos2011photonic}. One way to engineer the desired reflectivity at the design $\lambda_0$ wavelength is to analyze a single element (a cylindrical hole is chosen to ease the fabrication process) in the uniform waveguide and subsequently replicate it with an appropriate periodicity. The complex transmission coefficient of a single element $\rm T_{\rm M, single}$ can be obtained using a numerical simulation \cite{CST}, where a single cylindrical hole is placed in the waveguide and the structure is excited with the fundamental mode, $\rm M$. The purpose is to determine the phase shift introduced by the single element to the transmitted fundamental mode. 
	Consequently, with a careful selection of the distance $d$ between the hole centers (the periodicity), an optical mirror can be designed. 
	For the periodically placed holes to act as a mirror, the reflected waves of the individual holes (elements) must interfere constructively, while the forward scattered waves should interfere destructively.
	This is achieved if the \emph{electrical distance} $d/\lambda_{\rm M}$ between the hole centers --- where $\lambda_{\rm M}$ is the effective guided wavelength of the fundamental mode $\rm M$ at the design wavelength --- is an integer multiple of $\pi$, that is $n\cdot \pi$ with $n=1,2,3, \ldots$ The shortest physical length of the mirror is obtained for $n=1$.
	
	To achieve an electrical distance between the hole centers of $n\cdot \pi$, the phase delay encountered by a progressive wave traveling between two holes must be obtained. This phase delay consists of two parts: The first part is the phase shift due to propagation. For a harmonic time dependence given by exp$(i\omega t)$, a traveling wave undergoes a phase shift of $-2\pi d/\lambda_{\rm M}$, when propagating a (physical) distance $d$.
	The second part contributing to the phase delay is the phase shift the wave undergoes because it is scattered at the hole. This phase shift can be obtained from full-wave simulations of a single element (hole) and corresponds to the phase of the scattering parameter \cite{pozar2011microwave} $s_{21}$. Letting $\angle T_{\rm M,single} = \operatorname{arg}(s_{21})$, the phase of $s_{21}$ of a single hole embedded in the waveguide, the total phase delay is
	\begin{equation}
		-2\pi\frac{d}{\lambda_{\rm M}} + \angle T_{\rm M,single}
	\end{equation}
	
	\noindent which should be a negative multiple of $\pi$ to achieve an electrical distance of $n \cdot \pi$,
	\begin{equation}
		-n\cdot \pi = - \frac{2\pi}{\lambda_{\rm M}}d + \angle T_{\rm M,single}
	\end{equation}
	\noindent
	The minus sign on the left hand side appears because the time dependence was chosen to be exp$(i\omega t)$.
	Solving for $d$ yields an initial estimation of the necessary physical distance between the holes maximizing the reflectivity given by
	\begin{equation}
		d = \lambda_{\rm M} \left( \frac{n}{2} + \frac{\angle T_{\rm M,single}}{2\pi}   \right) \, . \label{delement}
	\end{equation}
	
	\begin{figure}[tb]
		\centering
		{\includegraphics[width=9cm]{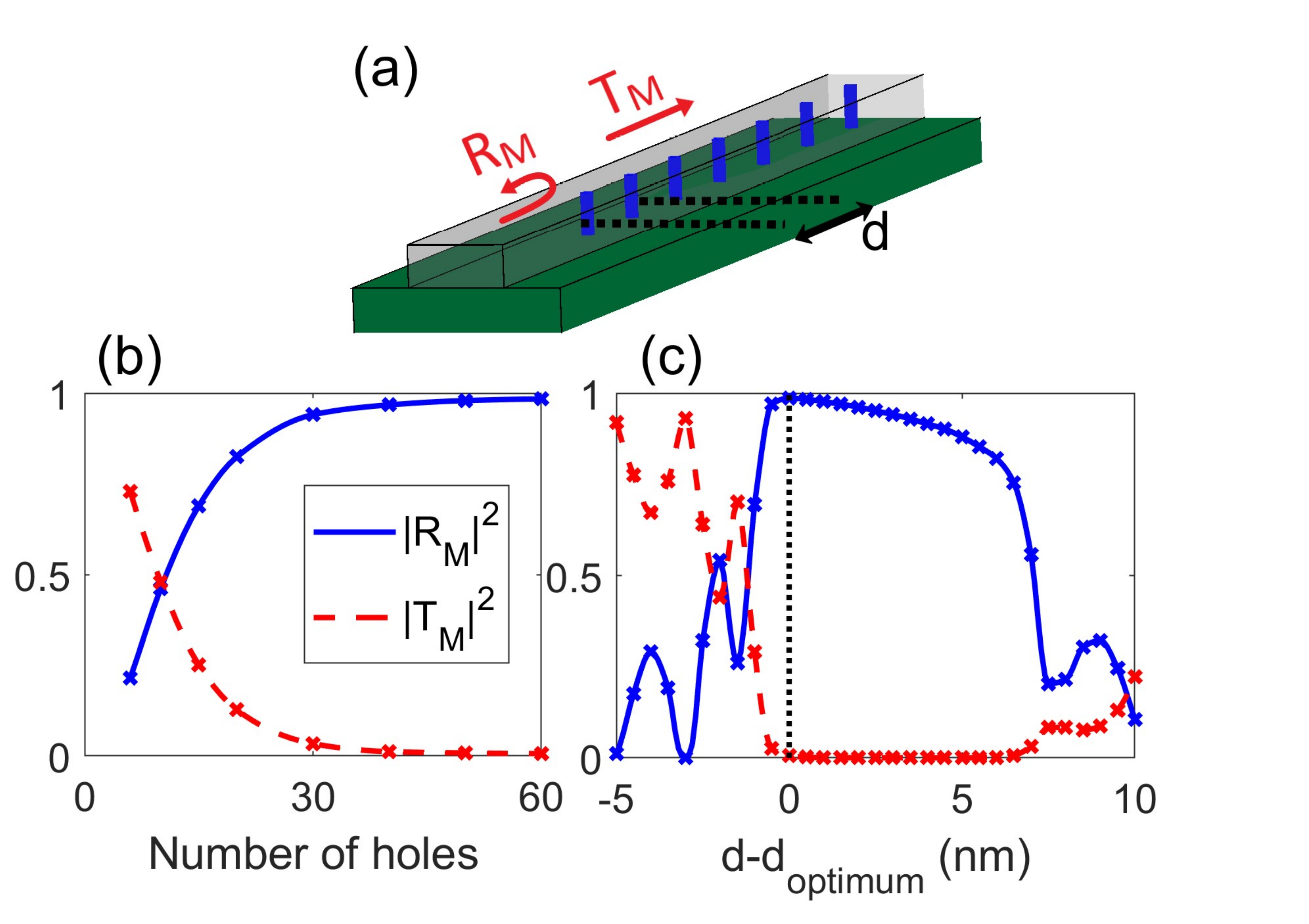}}
		\caption{Characteristics of the isolated photonic mirror. (a) Sketch of the mirror composed of holes with distance $d$ between the hole centers. (b) Power reflection and transmission coefficient for the fundamental mode as a function of the number of holes in the mirror for an optimal distance $d=d_{\rm optimum}=155$ nm between holes. (c) Tolerance of the power reflection and transmission coefficients for varying  the distance, $d$, for 30 holes.} \label{fig:mirror}
	\end{figure}
	
	We note that the hole diameter is a free design variable, and we now consider a mirror composed of $D$ = 50 nm diameter holes as shown in Fig.\ \ref{fig:mirror}(a). While a 50 nm diameter represents a challenge in the fabrication, it is not beyond state-of-the-art reactive ion etching. The choice of diameter is further discussed in Section \ref{sec:discussion}.
	
	The distance $d$ between the hole centers is now varied in the numerical simulation using the distance obtained from Eq.\ \eqref{delement} as the starting value, and an optimum distance $d_{\rm optimum}$ of 155 nm maximizing the reflectivity is identified. The power reflection and transmission coefficients for the fundamental mode as function of the numbers of holes are presented in Fig.\ \ref{fig:mirror}(b) for $d = d_{\rm optimum}$. While ten consecutive holes reflect half of the power, 60 holes reflect 0.99 at the design wavelength. To investigate manufacturing tolerances, we performed simulations for varying distance $d$ between the elements for a mirror with 30 holes, and the results are shown in Fig.\ \ref{fig:mirror}(c). It can be seen that variations in $d$ have significant influence on the reflectivity, where the reflection and transmission characteristics are more tolerant for distances larger than the optimal one. The asymmetry arises from the non-linear phase response of the single element as further discussed in Section \ref{sec:discussion}.
	Also, it should be noted that larger holes lead to significant scattering into background modes (i.e. anything different from the fundamental cavity mode, such as higher order and radiative modes), thus resulting in $|R_M|^2 + |T_M|^2 < 1$.
	
	\begin{figure}[tb]
		\centering
		{\includegraphics[width=0.9\linewidth]{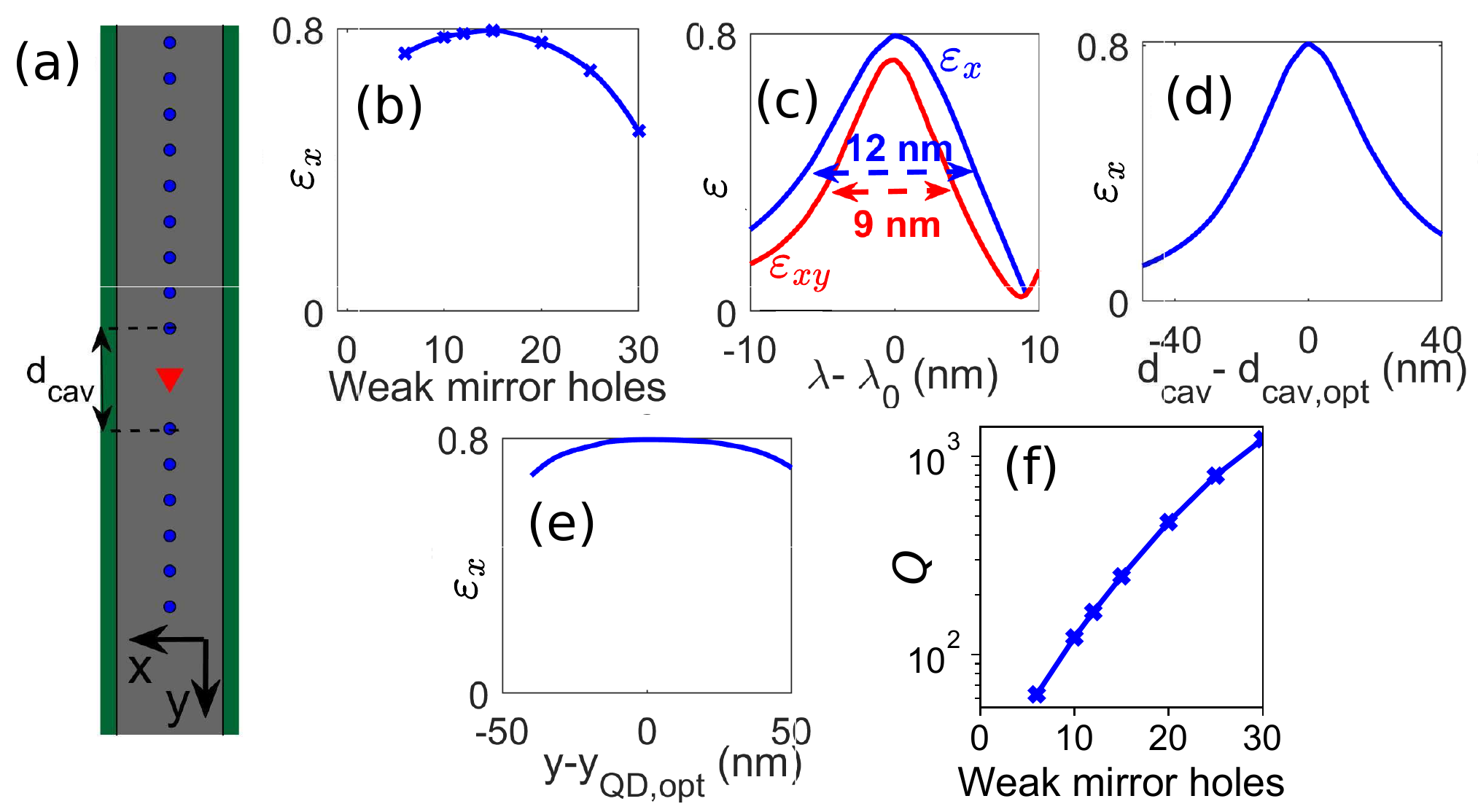}}
		\caption{
        Analysis of the cavity structure formed by two mirrors. (a) The top cross-sectional view of the cavity structure with the distance between the mirrors  denoted $d_{\rm cav}$ (hole center to hole center). (b) \change{$\varepsilon_{x}$} as a function of the number of the weak mirror holes for 60 holes in the strong mirror. (c) \change{$\varepsilon_{x}$} and \change{$\varepsilon_{xy}$} as a function of wavelength --- the horizontal lines denote the full-width at half maximum; the influence of (d) $d_{\rm cav}$ and (e) the QD location along the $y$ axis on \change{$\varepsilon_{x}$}, for a cavity with 15 (60) holes in the weak (strong) mirror. \NewChange{(f) Q factor as function of the number of weak mirror holes for 60 holes in the strong mirror.
		In all panels} we use $d_{\rm cav} = d_{\rm cav,opt} = 420$ nm and inter-hole distance $d=d_{\rm optimum}=155$ nm.}
		\label{fig:cavity}
	\end{figure}
	
	We proceed to form the cavity by combining a strong and a weak mirror. The strong mirror has simply the purpose of blocking the electromagnetic field from escaping, while the role of the weak mirror is to enhance the Purcell factor \cite{krasnok2015antenna} by partial reflection as well as to out-couple light to the desired output port. The top view of a cavity structure is shown in Fig.\ \ref{fig:cavity}(a). The cavity design parameter to be identified is now the distance $d_{\rm cav}$ between the strong and the weak mirror (center hole to center hole). Similarly to the distance between the elements, we can determine the distance between the mirrors. This time, our goal is to set the maximum field location at the source point to obtain the maximum Purcell enhancement to the fundamental mode.
	Therefore, the distance between the emitter and the weak (strong) mirror can be obtained as
	\begin{equation}
		d_{\rm{W(S)}}= \lambda_{\rm M}\bigg(\frac{n}{2}+\frac{\angle \rm R_{\rm M,W(S)}}{4\pi}\bigg), \label{dcav}
	\end{equation}  
	where $\angle \rm R_{\rm M,W(S)}$ is the reflection phase of the weak (strong) mirror. Consequently, our initial guess for the cavity length becomes $d_{\rm cav}=d_{\rm{W}}+d_{\rm{S}}$, and as before, we proceed to identify the optimum cavity length $d_{\rm cav,opt}$ of 420 nm using full-wave electromagnetic simulations starting from the initial guess.
	
\begin{figure}[tb]
		\centering
		{\includegraphics[width=13cm]{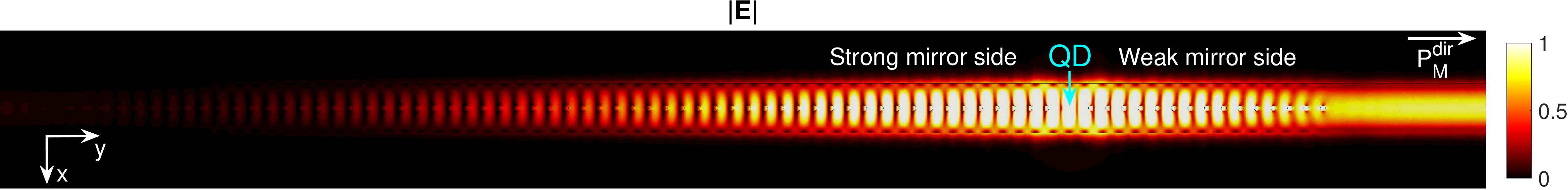}}
		\caption{Electric field distribution (absolute value) of the top cross-sectional view of the cavity ridge waveguide.} \label{fig:optEx}
	\end{figure}

	The \change{coupling efficiency $\varepsilon_{x}$}, calculated to the right of the weak mirror, is presented in Fig.\ \ref{fig:cavity}(b) as function of number of weak mirror holes with $d_{\rm cav} = d_{\rm cav,opt}$ and 60 holes in the strong mirror. We observe that the maximum \change{$\varepsilon_{x}$} is obtained for 15 holes in the weak mirror yielding a very attractive \change{$\varepsilon_{x}$} of 0.80. Taking into account the emission from the $y$ oriented dipole, a final \change{$\varepsilon_{xy}$} of 0.73 is obtained. 
	We thus obtain a performance comparable with previous work \cite{hepp2020purcell}, despite reducing the number of Bragg periods by a factor of 7.3.
	A top-cross-sectional view of the corresponding electric field generated by an $x$-oriented point dipole is presented in Fig.\ \ref{fig:optEx}. We observe that the field propagating to the left is effectively blocked by the strong mirror (60 holes), while the weak mirror (15 holes) efficiently out-couples the fundamental mode towards the right side with minimal scattering loss.
	
	The dependence of \change{$\varepsilon_{x}$ and $\varepsilon_{xy}$} for 15 weak mirror holes on the emission wavelength is shown in Fig.\ \ref{fig:cavity}(c), where we observe a similar asymmetric behavior as observed in Fig. \ref{fig:mirror}(c). Indeed, the relative effect of an increase in $\lambda$ (with fixed hole diameter $d$) is qualitatively similar to a decrease in $d$ (with fixed $\lambda$), leading to fast decrease in reflectivity as shown in Fig. \ref{fig:mirror}(c) and thereby decrease in \change{$\varepsilon_x$}.
	Even so, we observe a broad peak with a FWHM of $\sim$ 9 nm for \change{$\varepsilon_{xy}$} and $\sim$ 12 nm for \change{$\varepsilon_{x}$}. With such a broadband coupling, a spectral tuning mechanism \cite{hepp2020purcell} is not required, and this represents a major asset of our design. %\change{--- although this clearly comes at the price of a lower Purcell enhancement.}
	We now investigate the tolerance of our design to fabrication imperfection by presenting \change{$\varepsilon_{x}$} as function of $d_{\rm cav}$ and of the QD position $y_{\rm QD}$ along the $y$ axis in Figs.\ \ref{fig:cavity}(d) and (e), respectively. We observe that a deviation in $d_{\rm cav}$ from the optimum value $d_{\rm cav,opt}$ of $\sim$ 20 nm reduces the coupling by a factor of 2, whereas misalignment of the QD along the $y$ axis by 40 nm only reduces \change{$\varepsilon_{x}$} by $\sim$ 10 \%. The cavity geometry thus displays good tolerance towards fabrication imperfection, representing a second strong asset of our design.
	
	\NewChange{
	Finally, we present in Fig.\ \ref{fig:cavity}(f) the Q factor obtained from the complex eigenfrequency $\omega$ of the cavity mode as $Q$ = -Re($\omega$)/(2Im($\omega$)) \cite{Kristensen2012}.
	As expected, the $Q$ factor increases with the number of weak mirror holes. For the optimal design with 15 holes in the weak mirror, we have $Q$ = 248, a value more than one order of magnitude smaller than the one reported in Ref.\ \cite{hepp2020purcell} for a similar coupling efficiency.
	Furthermore, we estimate the mode volume $V$ from its relation with the $Q$ factor and the Purcell factor. We find a mode volume $V = 3.1 (\lambda/n)^3$ for the optimal design, where $n$ is the refractive index of GaAs. The dependence of $V$ on the number of holes in the weak mirror is marginal.}

	\section{Fabrication and measurements}\label{sec:fab}

\begin{figure}[tb]
		\centering
		{\includegraphics[width=13cm]{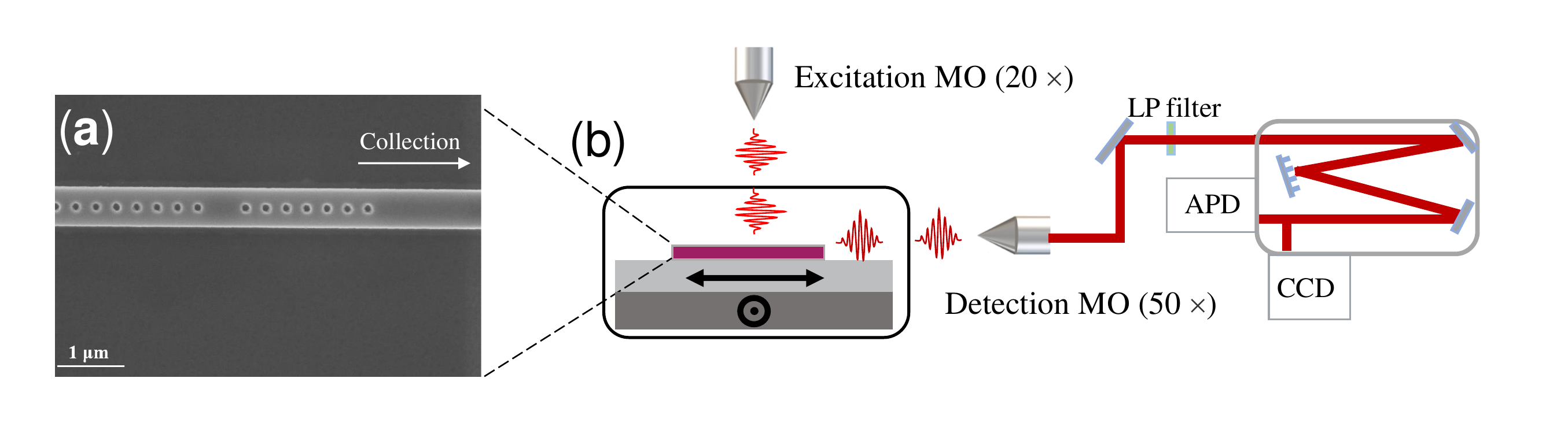}}
		\caption{(a) SEM image of a waveguide cavity structure with a single layer of InGaAs QDs as active layer. The diameter of the holes is around $120$ nm. The collection direction of emission is indicated by the white arrow. (b) Schematic view of $\mu$PL setup used to characterize the waveguide structures. The pump laser-excited QDs through a microscope objective (MO) from the top excitation path, and emission of waveguide-coupled QDs was collected by a second MO via the side collection path. The black arrow and dot-circle represent the directions of location control.} \label{fig:yuhui1}
	\end{figure}

	The InAs/GaAs QD sample used in this work was grown by metal-organic chemical vapor deposition (MOCVD) on a n-doped GaAs wafer. The ridge waveguide structures with periodic nanoholes forming an integrated cavity were patterned by standard electron beam lithography (EBL) and were subsequently etched by reactive ion etching in an inductively coupled plasma (ICP-RIE) with an etching depth of 250 nm between waveguide cavities. The increased etch depth compared to the 200 nm of the ridge waveguide design is partially due to a compensation for the shallower etch depth of small features like the holes of the cavities in the waveguide and an increased etch rate due to thermal heating. The epitaxial layer design is the same as described in Section \ref{sec:Design}.
	However, while the simulations show that a hole diameter $D = 50$ nm leads to very high coupling efficiency, minimal light scattering and the overall very low optical losses, we fabricated a structure with $D = 120$ nm due to limitations in the equipment available.
	Consequently, we opted to etch only 7 holes in the weak mirror (instead of 15) in order to extract more light from the cavity, as a higher hole diameter results in higher reflectivity and unwanted scattering.
	A scanning electron microscope (SEM) image of waveguide cavity structure is shown in Fig.\ \ref{fig:yuhui1}(a).
	For the later optical characterization, the sample was cleaved perpendicularly to the waveguide axis leading to 60 $\mu$m long waveguide sections each with an integrated nanohole cavity. We present the schematic view of our micro photoluminescence ($\mu$PL) setup in Fig.\ \ref{fig:yuhui1}(b). The sample is mounted in a liquid He flow cryostat operating at $20$ K. The cryostat contains a $x-y-z$ piezo-controlled translation stage enabling to precisely align the facet of the waveguide under study to the detection path. The QDs were optically pumped by a $680$ nm pulsed laser with a repetition rate of $f = 80$ MHz. The pump laser excited QDs through a microscope objective (NA $= 0.4$) from the top excitation path, and emission of waveguide-coupled QDs was collected by a second objective (NA $= 0.42$) via the side collection path. The collection direction is aligned to the side of the asymmetric cavity structure with a lower number of holes as indicated by a white arrow in Fig.\ \ref{fig:yuhui1}(a). After suppressing the laser scattering signal by a long-pass filter (RG $850$), the spontaneous emission of the waveguide cavity structure enters the monochromator and the spectrally resolved emission can either be collected by Si-based charge-coupled device (CCD) or a single photon counting module (SPCM) based on a Si avalanche photodiode (APD). Here the SPCM is used to perform time-resolved photoluminescence (TRPL) measurements.
	
\begin{figure}[tb]
		\centering
		{\includegraphics[width=\linewidth]{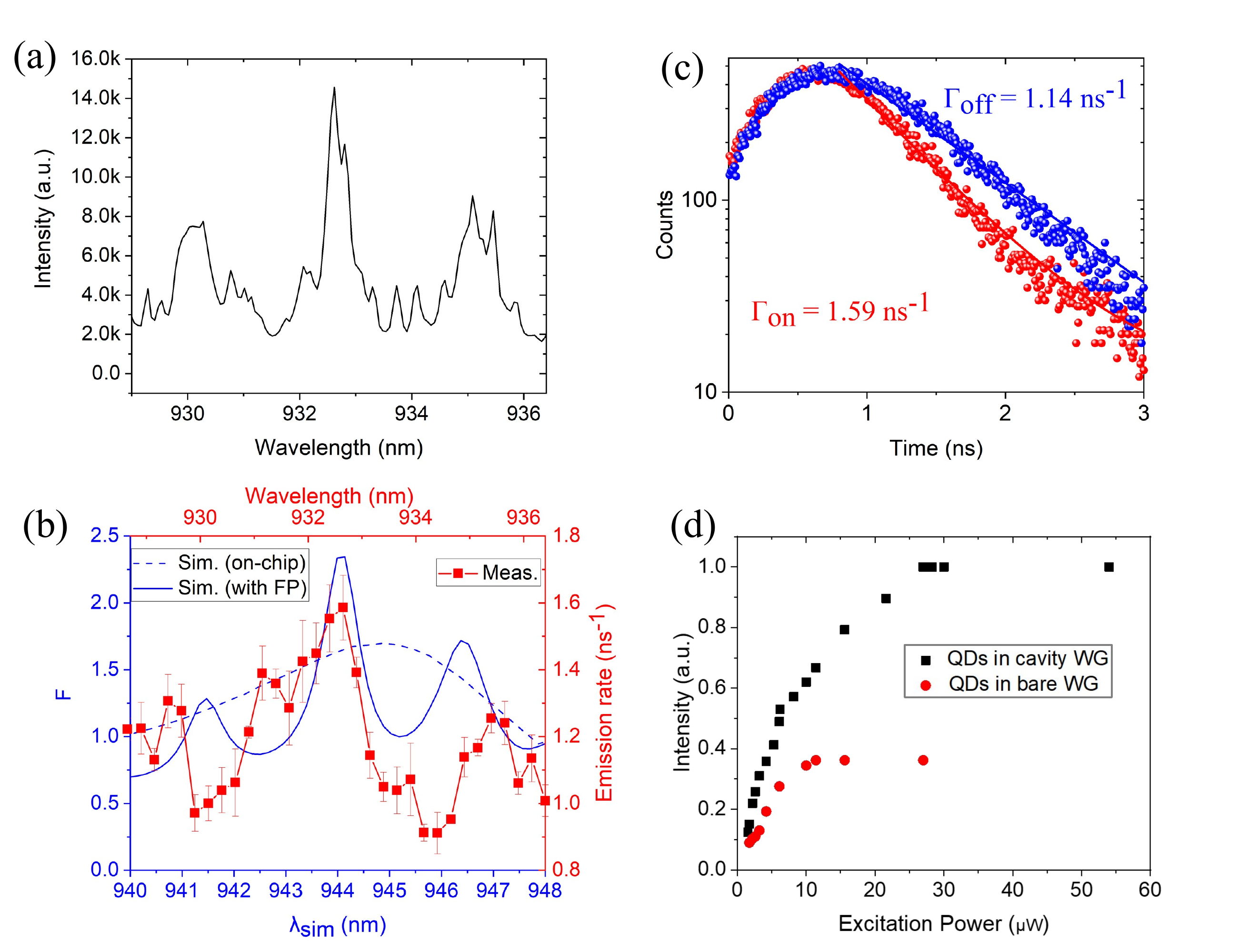}}
		\caption{Optical characterization of the waveguide cavity structure.
			(a) $\mu$PL spectrum of the waveguide cavity structure at an excitation power of 7 $\mu$W. Three emission peaks are observed at $929.8$ nm, $932.3$ nm and $934.8$ nm.
			(b) Spontaneous emission (SE) rate of the nanohole cavity’s resonance range (red). The panel also shows the simulated generalized Purcell factor $F$ of the structure with (solid blue curve) and without (dashed blue curve) truncation of the waveguide.
			Both experiment and theory exhibit the cavity enhancement due to light-matter interaction with Fabry-Pérot modes confined between the right mirror of the waveguide cavity structure and the right facet of the waveguide.
			(c) TRPL traces at resonance with an FP mode ($932.6$ nm, red curve) and off resonance ($\sim946$ nm, blue curve).
			(d) Power-dependent PL intensity measurement under above-band excitation ($780$ nm) by a cw laser.} \label{fig:yuhui2}
	\end{figure}

	First, we studied the emission properties of the waveguide cavity structure via pulsed above-band excitation, obtained from one cavity device containing an ensemble of QDs. We present the corresponding $\mu$PL spectrum (red curve at an excitation power of $7$ $\mu$W in Fig.\ \ref{fig:yuhui2}(a). Three discrete photonic resonances can be observed, decorated by single-QD lines of the active layer with a rather high QD density. These resonances are attributed to Fabry-Pérot (FP) modes confined by reflections at the right outer facet (with about 26~$\%$ reflectivity) of the relevant waveguide structure \cite{ba2012enhanced}. 
	These FP resonances with a free spectral range (FSR) of $\sim$ 2.5 nm modulate the cavity response.
	In fact, the observed FSR $= 2.5$ nm is in good agreement with the separation between the left mirror of the nanohole cavity (start point of strong mirror nanoholes) and the right facet to the waveguide structure with a separation of $\sim 40$ $\mu$m, together with the strong mirror’s effective penetration depth of $\sim$ 5.5 $\mu$m.
	
	For a more detailed understanding of the system’s dynamics, luminescence decay was measured additionally in a wavelength range of $928$ to $936$ nm, with a step size of $0.25$ nm, see red trace in Fig.\ \ref{fig:yuhui2}(b).
	Spectrally aligned with the FP peaks observed in Fig.\ \ref{fig:yuhui2}(a), we observe increased SE rates of $\Gamma_{\mathrm{on}}= (1.31 \pm 0.08)$ ns$^{-1}$, $(1.59 \pm 0.09)$ ns$^{-1}$, and $(1.25 \pm 0.04)$ ns$^{-1}$ for the resonance at $929.4$ nm, $932.6$ nm, and $935.3$ nm, respectively.
	Moreover, we show the simulation results for the generalized Purcell factor $F(\lambda)$ for the waveguide cavity structure [Fig.\ \ref{fig:yuhui2}(b), blue traces], where we have simulated a structure with identical hole diameter $D=120$ nm as the one fabricated in the experiment. While the dashed curve is for the on-chip calculation, where the truncation and reflections at the waveguide facet are not taken into account, the solid curve represents the simulation where the truncation of the waveguide is taken into account as an air gap after the waveguide end with a reflectivity of 0.26. As expected, the FP resonances are visible in the latter case, and they modulate the on-chip response in agreement with experiment. Comparing the simulation results with experiment, we observe a blue shift about $11$ nm, which we attribute to fabricated hole diameters being slightly larger than the design ($D = 120$ nm) diameter.
	
	To obtain further insight into the optical properties of the waveguide cavity structure and the underlying Purcell enhancement of emission we performed TRPL studies.
	Two exemplary TRPL traces are depicted in Fig.\ \ref{fig:yuhui2}(c). We recorded the PL decay curves at the resonance peak ($932.3$ nm, red trace) and out of resonance ($946$ nm, blue trace).
	In both cases, a slow rise of the signal is observed (raising time $\sim$1 ns) due to wide laser pulses being used for excitation, followed by an exponential decay.
	For the off-resonance case we determine a spontaneous emission (SE) rate of $\Gamma_{\mathrm{off}} = (1.14 \pm 0.14)$ ns$^{-1}$ ($\tau_{\mathrm{off}} = [0.87 \pm 0.10]$ ns), which is a typical value for InAs QDs in a ridge waveguide \cite{lund2008experimental,reithmaier2013chip}. In contrast, for the resonant case, we observe $\Gamma_{\mathrm{on}} = (1.59 \pm 0.09)$ ns$^{-1}$ ($\tau_{\mathrm{on}} = [0.63 \pm 0.04]$ ns) representing a $1.39$ times enhancement of SE due to enhanced light-matter interaction with the FP mode.
	
	To further confirm the cavity enhancement of the QD ensemble emission if compared with a bare waveguide, we performed power-dependent $\mu$PL measurements presented in Fig.\ \ref{fig:yuhui2}(d) of the waveguide cavity structure and of a bare waveguide with the same length. We observe that in the whole range of excitation powers, the integrated PL intensity of the resonance of the waveguide cavity structure is higher than that of the bare waveguide, and that the saturation intensity is about a factor of $2.5$ times higher in case of the waveguide cavity structure, indicating a higher total decay rate.
	While non-radiative decay channels may have been introduced in principle during the fabrication process (through, for instance, defect states coupled to the emitter), we observe that both the power required to reach saturation and the emission intensity at saturation seem to be $\sim2.5$ times higher in the presence of the cavity.
	This supports a scenario where most of the decays occur via photon emission in the cavity waveguide, indicating that the SE rate is indeed enhanced by the Purcell effect \cite{krasnok2015antenna}.
	
	\section{Discussion}\label{sec:discussion}
	
	We now discuss the strengths and drawbacks of our design procedure.
	Even though we are considering the low-index contrast GaAs/$\rm Al_{0.9}Ga_{0.1}As$ material platform, our cavity design allows for a coupling \change{$\varepsilon_{xy}$} of 0.73 thanks to significant Purcell enhancement of the light emission into the fundamental mode.
	Our design displays good tolerance towards spatial misalignment, and the large 9 nm linewidth means that Stark \cite{Somaschi2016} or strain-tuning \cite{hepp2020purcell} of the QD line to match a narrow resonance can be avoided.
	In our bare ridge waveguide optimization, we have assumed a QD placed at the center of the ridge waveguide. However, the geometry is asymmetric along the vertical axis, which shifts the antinode of the fundamental mode M towards the substrate as observed in the inset of Fig.\ \ref{fig:intro}. Also, to prevent the oxidation of AlGaAs cladding, an anti-oxidation layer may be introduced. This layer is typically the high-index waveguide material (GaAs), and its introduction pushes the antinode even further down. 
	The predicted performance in this work is thus slightly lower than what can be achieved by placing the QD exactly at the vertical field maximum. However, such an optimization is numerically demanding and is left for a follow-up work. 
	
	Our design procedure for identifying the hole periodicity is based on first-order reflection. The multiple reflections that arise between the holes are negligible for our small holes as the magnitude of the reflection coefficient for a mirror unit cell is small. However, for larger holes, the distances for the resonance condition require additional tuning to take into account higher-order reflection. Nevertheless, even in that case the method described in Section \ref{sec:Design} gives an excellent starting point for the optimization.
	
	\change{A major physical mechanism limiting the performance of the nanobeam SPS design is unwanted scattering at the cavity-DBR interfaces.}
	As the diameter of the holes increases, an increasing mismatch between the bare waveguide eigenmode and the DBR Bloch mode leads to increased scattering to higher order and radiative modes at the cavity-DBR interface limiting the maximum achievable \change{$\varepsilon_{x}$}. \change{We note that the fraction of unwanted scattering increases with the cavity Q factor, and thus simply increasing the Purcell factor proportional to Q/V does not guarantee an improved outcoupling efficiency. Rather, our design approach is based on an optimized bare waveguide geometry ensuring a low mode volume combined with dielectric screening of radiation modes \cite{stepanov2015quantum, hoehne2019numerical, Bleuse2011}, such that good coupling efficiency is obtained even for a moderate Q factor. To reduce the scattering,} 
	we have used a relatively small \change{hole} diameter ($D$ = 50 nm) for our optimized design. However, larger diameters are desirable not only in the fabrication, but also since larger holes lead to increased reflectivity for the periodic mirror unit cell and in turn a smaller mode volume. Here, one may consider a tapering strategy \cite{palamaru2001photonic,Lalanne2003} to suppress the scattering loss by gradually increasing the diameter of the holes surrounding the cavity.
	For instance, tapering was successfully implemented in SiN nanobeam cavities on a silica substrate \cite{Panettieri2016, Olthaus2020, Zhan2020}, demonstrating record-high Q factors for non-suspended structures.
	Also, it was shown that an elliptical hole shape gives rise to better reflectivity and larger photonic bandgap \cite{Quan2011, Panettieri2016, Zhan2020}.
	Again, we leave further optimization study of the cavity such as implementation of a taper and optimization of the hole shape to a follow-up work.
	
	The cavity design can be tuned spectrally by varying the cavity length $d_{\rm cav}$, the periodicity $d$ and the hole diameter. Increasing the distance $d$ between the elements leads to a red shift as the resonance condition now holds for longer wavelengths. In contrast, increasing the diameter of the holes leads to a blue shift: since the holes are vacuum ($\rm n_{\rm vacuum}<\rm n_{GaAs}$), larger holes require a longer waveguide section to compensate the phase difference, and the effect is thus similar to reducing the distance between the holes.
	
	Our design features spectral asymmetry due to the non-linearity of the phase response of a single hole. There is a direct correlation between the linearity of a single element and the symmetry of the spectral response of the overall design, and a circular grating element does not provide a linear phase response. Therefore the overall response is not exactly symmetrical, and the asymmetry becomes more pronounced with larger holes. Nevertheless, the broad linewidth of 9 nm is obtained in spite of this asymmetry.
	
	We believe that our design offers an attractive possibility for integrated photonics applications. 
	While GaAs waveguides have in general higher losses compared to silicon-based waveguides, high quality GaAs waveguides with losses as low as 1.4 dB/cm have been realized \cite{Pu2016}, thus paving the way towards the realization of on-chip, GaAs-based photonics circuits.
	An alternative approach for minimizing losses would be to scale this design in order to operate at telecom wavelengths. Here, a challenge would be to integrate quantum emitters at $\lambda \sim 1550$ nm (such as, for instance InAs/InP QDs) with Si-based platforms, which is currently subject of intense investigation \cite{Holewa2021}.
	\change{Furthermore,} hybrid integrated quantum photonic circuits may be proposed to take full advantage of the strengths of different materials platforms \cite{Elshaari2020}.
	
	\change{Finally, we notice that the mismatch between experiment and theory presented here is mainly related to the random integration of QDs inside the cavity. Better matching is expected in the future by deterministically integrating single QDs at the field antinode of the cavity using in-situ electron beam lithography \cite{schnauber2018deterministic}.}
	%For instance, high atom-cavity cooperativity has been obtained experimentally by deterministic implantation of SiV centers in diamond nanobeam cavities \cite{Sipahigil2016, Nguyen2019}}

\section{Conclusion} \label{sec:conc}

In conclusion, we have designed a ridge waveguide nanobeam cavity in the GaAs/$\rm Al_{0.9}Ga_{0.1}As$ material system featuring a broadband \change{coupling efficiency $\varepsilon_{xy} \approx 0.73$} for a QD with in-plane dipole orientation. The design employs asymmetrically placed nanobeam mirrors formed by circular-cylindrical holes with diameter $D$ = 50 nm as Bragg gratings. The large coupling is obtained over a 9 nm linewidth, which allows for spectral alignment without the need for sophisticated tuning mechanisms. 
Furthermore, to demonstrate the cavity effect on the light emission, we have fabricated and characterized a fabrication-friendly ($D$ = 120 nm) ridge waveguide nanobeam cavity. The $\mu$PL measurements provide clear evidence of the cavity-induced Purcell enhancement in very good agreement with the theory. 
Finally, we have sketched the path towards further increasing the coupling efficiency using this platform. 
Our approach represents a significant step forward in performance within ridge waveguide single-photon source designs based on low-refractive-index-contrast materials.

\begin{backmatter}
	\bmsection{Funding}
	The authors acknowledge support from Villum Fonden (VKR Center of Excellence NATEC-II, grant 8692), from the European Research Council (ERC-CoG "UNITY", grant 865230), from the European Union’s Horizon 2020 Research and Innovation Programme under the Marie Sklodowska-Curie Grant Agreement No. 861097 and from the Independent Research Fund Denmark (grant DFF-9041-00046B). 
	
	\bmsection{Disclosures}
	The authors declare no conflicts of interest.
	
	\bmsection{Data Availability Statement}
Data underlying the results presented in this paper are not publicly available at this time but may be obtained from the authors upon reasonable request.
\end{backmatter}
%%%%%%%%%% If using BibTeX:
\bibliography{sample}

\end{document}